# Operation of a quantum dot in the finite-state machine mode: single-electron dynamic memory

M. V. Klymenko,[1] M. Klein,[2] R. D. Levine,[2,3] F. Remacle[1,2a)]

[1] Department of Chemistry, University of Liège, B4000 Liège, Belgium

[2] The Fritz Haber Center for Molecular Dynamics and the Institute of Chemistry, The Hebrew University of Jerusalem, Jerusalem 91904, Israel

[3] Crump Institute for Molecular Imaging and Department of Molecular and Medical Pharmacology, David Geffen School of Medicine and Department of Chemistry and Biochemistry, University of California, Los Angeles, CA 90095

Abstract

A single electron dynamic memory is designed based on the non-equilibrium dynamics of charge states in electrostatically-defined metallic quantum dots. Using the orthodox theory for computing the transfer rates and a master equation, we model the dynamical response of devices consisting of a charge sensor coupled to either a single and or a double quantum dot subjected to a pulsed gate voltage. We show that transition rates between charge states in metallic quantum dots are characterized by an asymmetry that can be controlled by the gate voltage. This effect is more pronounced when the switching between charge states corresponds to a Markovian process involving electron transport through a chain of several quantum dots. By simulating the dynamics of electron transport we demonstrate that the quantum box operates as a finite-state machine that can be addressed by choosing suitable shapes and switching rates of the gate pulses. We further show that writing times in the ns range and retention memory times six orders of magnitude longer, in the ms range, can be achieved on the double quantum dot system using experimentally feasible parameters thereby demonstrating that the device can operate as a dynamic single electron memory.

*Index Terms*—quantum dots, single electron memory, single electron transistor, DRAM cell, charge sensor

---

[a)] Corresponding author, email : fremacle@ulg.ac.be

# I. INTRODUCTION

Electrostatically-defined quantum dot (QD) electronic devices are promising elements for quantum computing systems that are compatible with silicon VLSI technology.[1] Moreover, due to their low energy consumption and small sizes, lateral QDs have also found applications in classical computing systems such as binary and multivalued logic gates,[2] and single-electron memory.[3, 4] In this paper, we further extend their potential range of application and demonstrate the ability of electrostatically-defined QD's to operate as a finite state machine [5, 6] performing sequential logic operations or providing single electron dynamic memory.

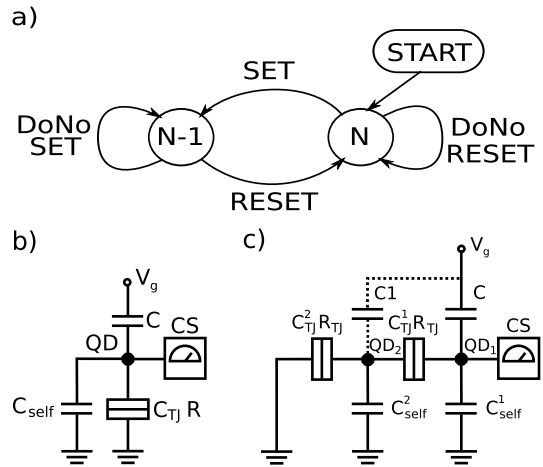

Fig. 1 a) The state diagram of a set-reset finite state machine (DoNo stand for the Do Nothing operation). b) and c) the circuit diagrams of the QD-CS (Quantum Dot - Charge Sensor) devices operating in the finite-state machine mode, b) is a single QD device and c) is a double QD device coupled by a tunnel junction.

In classical electronics, designing sequential logic devices is closely related to the concept of the feedback.[6] However, confined quantum systems inherently offer a set of states that can provide multistable dynamics without a feedback. In this paper we report the design of a dynamic single electron memory based on electron charge states of QD's. Existing single-electron memories on QDs[3, 4] are based on floating gates and multiple tunnel junctions that exhibit a hysteresis in their transfer characteristics. [3, 7] In these devices, the memory is volatile and switching between the logic levels is accompanied by changing the QD occupation by 5-10 electrons. In this work, in order to significantly decrease the energy consumption, we discuss the regime of operation where the logic states of the memory differ by QD occupations of just a single electron. One of the most efficient ways to switch between charge states is based on the turnstile single-electron pump.[8] However, the turnstile pump requires an additional gate electrode and a complicated timing of the applied gate pulses that makes device addressing technologically more difficult.

In this paper we aim to get a specific regime of the QD operation that can be described as a dynamical finite state machine (FSM),[5, 6] whose state diagram is shown in Fig. 1a. The two states of the FSM are encoded by two charge configurations of the QD with number of electrons $N$ and $N$-1 respectively. Initially the QD is chosen to be in the state $N$. The logic operation SET of the FSM leads to the charge state $N$-1. The RESET logic operation brings the system back to the state $N$. Such a FSM is the simplest device exhibiting a memory effect: if no action is applied (the Do Nothing (DoNo) operation in the diagram of Fig. 1a) the device should hold its current state. If a transition does occur during a DoNo interval, the system generates a bit error. The memory storage time is therefore determined by how long the QD keeps its current charge state, either $N$ or $N$-1, compared to the time it takes to implement a SET or a RESET operation. We show that the storage time can be maximized by a proper choice of the duration, amplitudes and switching rates of voltage pulses applied to the gate electrode of the QD. We also compute the probabilities of being in a given charge state which allows to estimate the bit error rate. The development of a short-term dynamic single-electron memory is motivated by its potential applications: the implementation of DRAM memory cells,[8] of a single-electron finite-state machine[5] and the physical realization of short-time synaptic plasticity in neuromorphic electronics.[9] The finite lifetime of a charge state of the QD device is the determining factor in the retention of the state. One of our objectives is to maximize this time. Note that even when the probability of retaining the state is not equal to one, but high enough, the device is of interest for approximate computing systems.[10]

Since the QD device operates in the single electron limit, the energy dissipation caused by the SET and RESET operations is closer to the Launder limit than ordinary CMOS based ones.[11-13] The main contribution to the energy consumption comes from the tunneling process through the resistive junction to the electron reservoir, as shown in Figs. 1b and 1c and from probing the charge state.

## II. QD-CS DEVICES AND THEIR PARAMETERS

The devices investigated are shown in Fig. 1b and Fig. 1c. They consist of a single or double QDs respectively, coupled to a charge sensor (CS) that provides a non-invasive probe of the charge state. In what follows we use the abbreviation QD-CS for the device. Two kinds of real-time charge sensors are currently available: one is based on quantum point contacts [14] and the other one on single electron transistors.[15] Here we chose the latter because sensors based on single electron transistors are characterized by a faster measurement time (~ 2 μs at the signal to noise ratio 9.0 in Ref.[15]).

The QD-CS devices operate as a single-electron box[11] since they have a single gate electrode and are connected to an electron reservoir by means of a single tunnel junction. In Fig. 1b and Fig. 1c, the gate electrode with voltage $V_g$ is electrostatically coupled to the system with the coupling capacitance $C$. The capacitances $C_{TJ}$ describe the electrostatic coupling through tunnel junctions, while the capacitance $C_{self}$ models the electrostatic influence of the environment and all top-gates that are not explicitly included in the model.

In the single-dot device, the capacitances $C_{TJ}$ and $C_{self}$ equal $300 \cdot q$ F, where $q$ is the electron charge, $q=|e|$. The gate coupling capacitance, $C$, is $100 \cdot q$ F. These values are chosen to be experimentally realistic, so that the total dot capacitance $C_\Sigma = 700q$ F and the QD charging energy equals 1.4 meV.

In QD-CS device shown in Fig. 1c, two QDs are arranged in a chain such that the gate electrode is attached to only one QD (QD1, in Fig. 1c). QD1 is coupled to the second QD (QD2, in Fig.1c) through a tunnel junction. QD2 is also coupled to an electron reservoir. QD1, that is attached to the charge sensor, has the same self-capacitance and charging energy as in the single-dot device: $C_{self}^1 = 300 \cdot q$ and $C_{TJ}^1 = 100 \cdot q$. The capacitances of QD2 are slightly larger with parameters $C_{self}^2 = 280 \cdot q$ F and $C_{TJ}^2 = 300 \cdot q$ F. The dots are not equivalent in order to be able to define the operational point in the stability diagram when the gate voltage is zero (see discussion on that point in the next section). Due to the small size of the system we also take into account parasitic electrostatic coupling between the gate electrode and QD2 that is modeled by the capacitance $C1 = 30 \cdot q$ F. The value of the tunnel junction resistances, $R_{TJ}$, taken to be 2 $M\Omega$ in all cases, is compatible with electron transfer dynamics on a nanosecond time-scale and approximately 80 times larger than the minimal tunnel resistance $h/2q^2$.

Besides electrostatically-defined lateral QDs, the circuit diagrams shown in Fig. 1b and 1c can be also implemented in a single nano-scaled FinFET transistor as has been demonstrated recently in ref.[13]. In this case, the QD and charge sensor are formed in corner conducting channels of a transistor's fin as a superposition of electrostatic potentials of the FinFET confinement potential, impurity potentials and gate electrode potential.

## III. SIMULATION OF TRANSIENT PROCESSES IN SINGLE-ELECTRON BOX

### A. Model

The operation of the QD-CS devices is simulated using the capacitance model [16, 17] for the spectrum of charge states. A master equation [18] together with the orthodox model are used for computing the tunneling rates [19, 20] and the dynamical response of the system. This model, while requiring only modest computational resources, allows reproducing accurately the dynamical response of the charge state to variations of the voltage levels at all electrodes.

In the framework of the capacitance model, the stable charge configurations can be found by minimizing the electrostatic free energy $F$ [16] namely the work of charging:

$$F = \frac{1}{2\mathbf{C}}\left(e\mathbf{N} - \mathbf{C}_{DV}\mathbf{V}^T\right)^2 \tag{1}$$

where $\mathbf{V}$ is the vector whose components are the voltages applied at all voltage nodes, $\mathbf{N}$ is the vector whose components are numbers of electrons on the n QDs that make the device, $\mathbf{C}_{DV}$ is the matrix of all capacitances describing the electrostatic coupling between the n QDs and the electrodes.

The values entering Eq. (1) read for the single-dot ($n=1$) device:

$$\mathbf{N} = N, \qquad \mathbf{C} = C + C_{TJ} + C_{self},$$

$$\mathbf{V} = \begin{pmatrix} 0 \\ V_r \\ V_g \end{pmatrix}, \qquad \mathbf{C}_{DV} = \begin{pmatrix} C_{self} & C_{TJ} & C \end{pmatrix},$$

and for the double-dot ($n=2$) device:

$$\mathbf{N} = \begin{pmatrix} N_1 \\ N_2 \end{pmatrix}, \qquad \mathbf{C} = \begin{pmatrix} C_\Sigma^1 & -C_{TJ}^1 \\ -C_{TJ}^1 & C_\Sigma^2 \end{pmatrix},$$

$$C_\Sigma^1 = C_{self}^1 + C_{TJ}^1 + C,$$

$$C_\Sigma^2 = C_{self}^2 + C_{TJ}^1 + C_{TJ}^2$$

$$\mathbf{V} = \begin{pmatrix} 0 \\ 0 \\ V_r \\ V_g \end{pmatrix}, \qquad \mathbf{C}_{DV} = -\begin{pmatrix} C_{self}^1 & 0 & 0 & C \\ 0 & C_{self}^2 & C_{TJ}^2 & C1 \end{pmatrix}.$$

The voltage $V_r$ is the potential of the electron reservoir. For the actual device operation it will be grounded. However, to analyze the time of relaxation processes in QDs, we allow it to be changed.

The stable charge configuration minimizes the free energy and is defined mathematically as $\mathbf{N}_s = \arg\min_{\mathbf{N} \in Z^n} F(\mathbf{N})$,[21] where $Z^n$ is the *n*-dimensional field of the positive integers including zero. The probability of finding the system in the charge configuration labeled $N$ reads $p_N$. The time evolution of $p_N$ is governed by the master equation[18]:

$$\frac{dp_N}{dt} = \Gamma_{N,N+1} p_{N+1} + \Gamma_{N,N-1} p_{N-1} - (\Gamma_{N+1,N} + \Gamma_{N-1,N}) p_N \qquad (2)$$

where $\Gamma_{j,i}$ is the transition rate from the charge state *j* to the charge state *i*. Eq. (2) implies that the occupancies on the dots can be changed by accepting or removing a single electron only. The system (2) represents an infinite chain of coupled equations involving all possible charge states. We truncate at three nearest to *N* charge states using the procedure proposed in Ref. [18].

Using the orthodox model,[19, 20] the tunneling rate governing the relaxation transitions between two charge states *i* and *j* is:

$$\Gamma_{i \to j} = -\Delta E_{i,j} \Big/ q^2 R_{TJ} \left[ 1 - \exp\left(\Delta E_{i,j} / kT\right) \right] \qquad (3)$$

where $\Delta E_{j,i}$ is the energy difference between the free energies of states *j* and *i*, *k* the Boltzmann constant and *T* the electronic temperature.

This model for the transition rates is valid for metallic QDs, where the spacing between the QD energy levels at a constant number of electrons is less than *kT*, while the spacing between charge states must be larger than *kT*.[19, 20] The main property of this expression which we will exploit later is that the transition has a non-negligible value only when $\Delta E_{j,i}$ is negative and it is negligible otherwise as shown in Fig. 2a.

For the double-dot QD-CS device, we are only interested in the occupation probability for QD1, that is attached to the charge sensor. It is computed as $p(N_1) = \sum_{N_2} p(N_1, N_2)$ where $p(N_1, N_2)$ is the probability of being in the state *(N₁, N₂)*.

In order to find the dynamical response of the QD system, we first compute the free energy and transition rates (Eqs. (3)-(5)) at a given voltage level at the gate electrode. Then we solve the system of master equation (2) using a fourth-order Runge-Kutta integrator.

B. **Switching between charge states in the single quantum dot device**

Coulomb blockade results in a stability diagram that characterizes the QD. The diagram delineates stability regions corresponding to a set of gate voltages for which the electrical current through the QD is blocked and the QD has a well-defined charge configuration. However, an instant transition from one stability region to another does not imply an infinitely fast change of the charge state: this process takes a time defined by transition rates, Eq. (3).

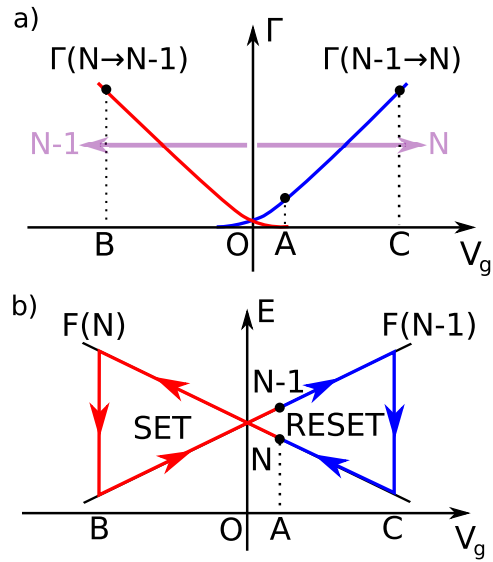

Fig. 2 a) The transition rates of the single-electron box, b) the energy diagram in the vicinity of the degeneracy point between charge configurations *N-1* and *N*. The arrows in panel b) show the changes of the electron energy as a result of two kinds of applied voltage pulses: the pulse SET is characterized by voltage levels A and B and crosses the degeneracy point O. Both voltage levels of the RESET pulse remain within the same stability region diagram. The SET pulse brings the system from charge state *N-1* to charge state *N*, while the RESET pulse brings it back to state *N*.

In Fig. 2a, we plot a sketch of the transition rates, defined by Eq. (3), between two charge configurations, $\Gamma_{N \to N-1}$ and $\Gamma_{N-1 \to N}$ near the degeneracy point between two stability regions. The degeneracy point is at $V_g = 0$. The energy diagram, shown in Fig. 2b, illustrates that, at the degeneracy point, two charge configurations have equal free energies, $F(N-1) = F(N)$ and they are linear in the vicinity of that point. Although all points of a particular stability region are characterized by the same charge configuration, they do not have equal energies and transition rates.

Recently, the energy dissipation in such a system has been studied in detail by measuring the rf ac response of the single-electron box on applied gate voltage oscillations.[11, 13] In this case, the energy dissipation, resulting in so-called Sisyphus resistance, strongly depends on the bias point of the rf signal relative to the degeneracy point.[13]

In this work, using the model presented in III.A, we study the charge state relaxation process resulting from biasing the system out of thermal equilibrium by an applied gate voltage pulse. The voltage pulse is represented by a linear combination of hyperbolic tangents:

$$V_g(t) = \frac{1}{2}(V_1 - V_0)\{\tanh[\alpha(t - t_1)] - \tanh[\alpha(t - t_2)]\} + V_0 \qquad (6)$$

where $1/\alpha$ is a pulse rising time. The pulse is characterized by two levels, $V_0$ and $V_1$, and the pulse width $t_2 - t_1$. The rising and decaying times of the pulses, $1/\alpha$, equals 0.01 ns.

First we apply the SET pulse symmetrically relative to the degeneracy point $O$. The initial point $V_0 = A$, shown in Fig. 2, is at a gate voltage of 5 mV so that the system occupies the center of the stability region with the number of electrons $N$. From this point we shift the gate voltage to $V_1 = -5$ mV, point B, that lies in the center of the stability region for the charge configuration $N-1$. To actually bring the system to the stable charge $N-1$ one needs to stay some time at that point until the relaxation process, which drives the system to its energy minimum, has been finalized, since the gate voltage switching rate was faster than the relaxation rate. The time needed for the system to relax determines the pulse width after which we switch back to point A, reaching a new non-equilibrium charge configuration. Since the rate of switching, $\alpha$ is much faster than the relaxation rates, the switching between charge states is a diabatic process and is accompanied by energy dissipation that is illustrated in Fig. 2 b by vertical arrows.

The corresponding changes of the occupation probability are shown in Fig. 3 (solid line). Switching the gate voltage back and forth is characterized by approximately equal rates of the transient processes. This situation dramatically changes when we shift both low and high voltage levels, $V_0$ and $V_1$, close to the borders of the stability regions by applying the pulse asymmetrically relative to the degeneracy point $O$. In this case, the probability for $N$-$1$ at the left front of the pulse rises faster, while its decay at the right front takes a significantly longer time (see Fig. 3, dashed and dotted lines). The closer we get to the degeneracy point, the larger the asymmetry of the transition rates. However, for applications, the voltage bias between the zero operating point $A$ and the degeneracy point must be chosen carefully, since at small difference the signal from the charge sensor bears a high level of the burst noise resulting from random jumps between charge states $N$ and $N$-$1$. Moreover, the asymmetry in the transition rates is vanishing when the temperature increases (the data shown in Fig. 3 have been computed for 30 mK).

The non-stable charge state $N$-$1$ at point $A$ relaxes to the state $N$. This process may be speed up applying a RESET pulse with opposite polarity that brings the system far from the point $O$ (see Fig. 2 b) where the relaxation rate to state $N$ is larger.

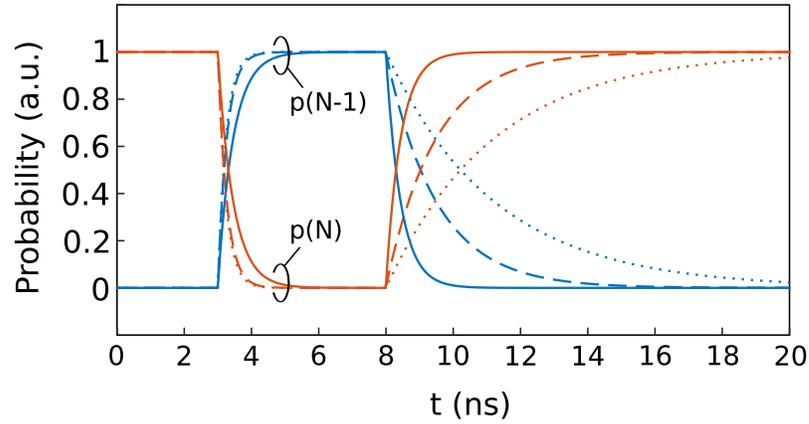

Fig. 3 Time-dependence of the electron occupation probability of $N$-$1$ and $N$ charge states resulting from applying the pulse of the gate voltage. The solid lines correspond to the pulse for which $V_0$=5 mV and $V_1$=-5 mV are localized at the centers of stability regions. The dashed and dots lines stand for the case when the low and high levels of the pulse are close to the edges of the stability regions ($V_0$=1.4 mV, $V_1$=8.6 mV for the dashed lines and $V_0$=-0.8 mV, $V_1$=9.4 mV for the dotted lines).

## C. Markovian electron relaxations in the double-quantum-dot system

The asymmetry between transition rates discussed in section III.B is exploited for operating a single electron finite-state machine by optimizing the ratio between the switching and the retention times of a state. To achieve this goal, we use a QD-CS device with two quantum dots forming a chain (see Fig. 1c) to enhance the asymmetry in the backward and forward switching rates between two charge states.

In a several electron system, bringing a chain of quantum dots from one stable state to another is accompanied by a large finite number of transitions through transient states as illustrated schematically on the state transition diagram plotted in Fig. 4. How complicated this process is depends on the applied bias and the number of quantum dots in the chain. For instance, the transition from the charge state (0, 0) to the state (1, 1) in the double-dot system involves two transient states (see Fig. 4 b). Compared to a single dot, the relaxation involves more transient states. Each addition of a QD to the chain increases the number of possible paths between two states, see figure 4c. The electron transport in a chain of metallic quantum dots, known as a multi-tunnel-junction, is characterized by many interesting processes including doubling of conductance peaks and hysteresis in the current-voltage characteristic[3] that have been intensively studied before. However, to the best of our knowledge, the relaxations processes in such systems have not been studied in details yet.

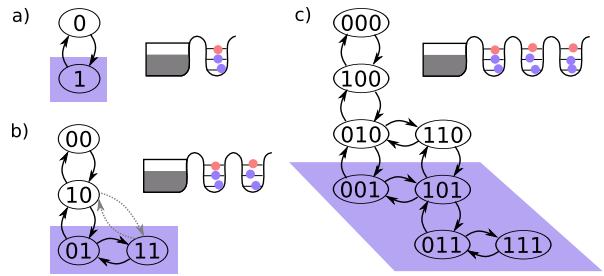

Figure 4 The state transition diagrams for a chain formed by a) a single, b) two and c) three quantum dots. Sequential tunneling is shown in black full lines and cotunneling in dotted grey lines.

The sequential tunneling represented on the diagrams of Fig. 4 is a stochastic process which can be modeled accurately by a continuous-time Markov chain neglecting the non-Markovian electron transport. In this work we also neglect the co-tunneling events (shown in dotted grey lines in fig 4b) due to low temperature and small voltage biases. The Markov process can be successfully modeled using a master equation of the form of Eq. (2) that can be derived from the Chapman–Kolmogorov equation.[22]

Here we consider the relaxation processes towards thermodynamical equilibrium in the chain of two coupled QDs whose parameters are described in the Section II. The stability diagram of the two coupled dots in terms of the gate voltage and the voltage of the electron reservoir is shown in Fig. 5. When the voltages at the gate electrode and reservoir are equal to zero, the system occupies the upper-left corner of the stability region for the charge configuration *(N, N-1)*. The left number characterizes the occupation of the dot attached to the charge sensor, QD1, while the right number corresponds to the dot connected to the electron reservoir, QD2 in Fig. 1c.

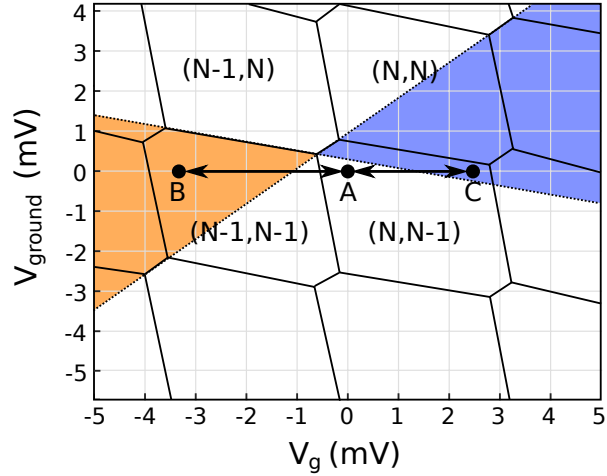

Figure 5 The stability diagram of the double-quantum-dot system. The areas shaded in orange on the left and in blue on the right correspond to the regions of voltages where the transition rates $\Gamma[(N,N-1)\to(N-1,N)]$ and $\Gamma[(N-1,N)\to(N-1,N-1)]$ (left) and $\Gamma[(N-1,N-1)\to(N-1,N)]$ and $\Gamma[(N-1,N)\to(N,N-1)]$ (right) are not zero.

We apply the same operating principle as has been discussed in section III.B. From point A in figure 5 where the stable state *(N,N-1)* the gate voltage is diabatically shifted to point *B* to a new stable charge state *(N-1,N-1)* and diabatically switched back after several nanoseconds. The direct transition from state *(N,N-1)* to that state *(N-1,N-1)* is forbidden since the electron can escape from the dot QD1 only through the dot QD2, so the number of electrons at the dot QD2 should be changed accordingly. However, point B is localized in the region where both the transition rates $\Gamma[(N,N-1)\to(N-1,N)]$ and $\Gamma[(N-1,N)\to(N-1,N-1)]$ are not zero. Therefore the state $(N-1,N-1)$ can be reached through the state $(N-1,N)$. The values of the corresponding transition rates as a function of the gate and reservoir voltages are shown in Fig. 6a and b respectively. The transition rate $\Gamma[(N,N-1)\to(N-1,N)]$ shown in Fig. 6a is characterized by a threshold that goes along the edge separating the stability regions of the charge state $(N,N-1)$ and $(N-1,N)$ (see Fig. 5). In turn, the edge between the charge states $(N-1,N)$ and $(N-1,N-1)$ determines the threshold for the transition rate $\Gamma[(N-1,N)\to(N-1,N-1)]$. These two edges form a zone in the stability region (shaded region in orange in Fig. 5 on the left) determining the voltage level of pulses for which the transition from the charge state $(N,N-1)$ to the state $(N-1,N-1)$ is possible. After being in the point B a certain time the system has relaxed to the state $(N-1,N-1)$. Switching the gate voltage back to the point A, the relaxation process is governed by the transitions rates $\Gamma[(N-1,N-1)\to(N-1,N)]$ and $\Gamma[(N-1,N) \rightarrow (N, N-1)]$ shown in Fig. 6 c and d. The thresholds for these transition rates form a cone shown as an area shaded in blue on the right side in Fig. 5.

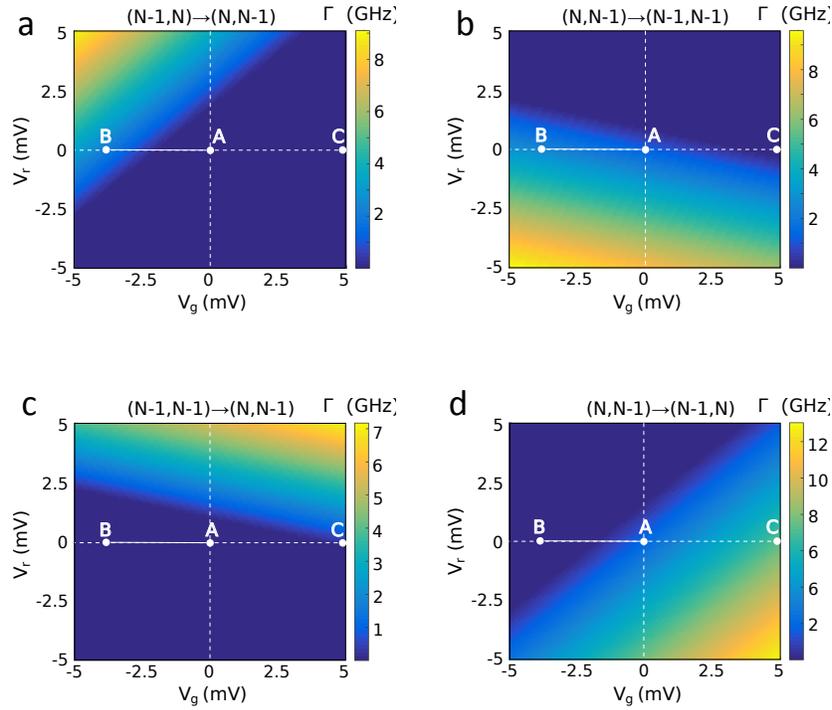

Figure 6 Computed transition rates for the transitions between charge states: a) $(N, N-1) \rightarrow (N-1, N)$, b) $(N-1, N) \rightarrow (N-1, N-1)$, c) $(N-1, N-1) \rightarrow (N-1, N)$ and d) $(N-1, N) \rightarrow (N, N-1)$ as a function of the gate and reservoir voltages

The results of the transition rates computations (see Fig. 6 c) evidence that the point $A$ is not localized in that region. The relaxation rate $\Gamma[(N-1, N-1) \rightarrow (N-1, N)]$ is very small and non zero due to finite temperature. Therefore, the system remains in the state $(N-1, N-1)$ for a long time. However, going farther to the point $C$ will bring the system back to the state $(N, N-1)$.

The modeling of the system dynamics using the master equation (Eq. (2)) allows to estimate the exact characteristic times of the processes described above. The occupation probabilities resulting from the solution of the master equation for a sequence of two voltage pulses: the first pulse goes through the points *A-B-A* on the stability diagram (the SET pulse), the other, applied 5 µs after, goes through points *A-C-A* (the RESET pulse) are shown in Fig. 7a. The results plotted in panel b) evidence that the relaxation $(N, N-1) \rightarrow (N-1, N-1)$ involves going through the transient state *(N-1,N)* with a high enough probability as has been discussed above. Also we observe an extremely high asymmetry in the relaxation times: the relaxation time for the transition $(N, N-1) \rightarrow (N-1, N-1)$ at the point B is less than 4 ns, while the relaxation time for the transition $(N-1, N-1) \rightarrow (N, N-1)$ at the point A is 2.3 ms at the temperature of 3 K (see fig. 7c). The two relaxation times differ by 6 orders of magnitude.

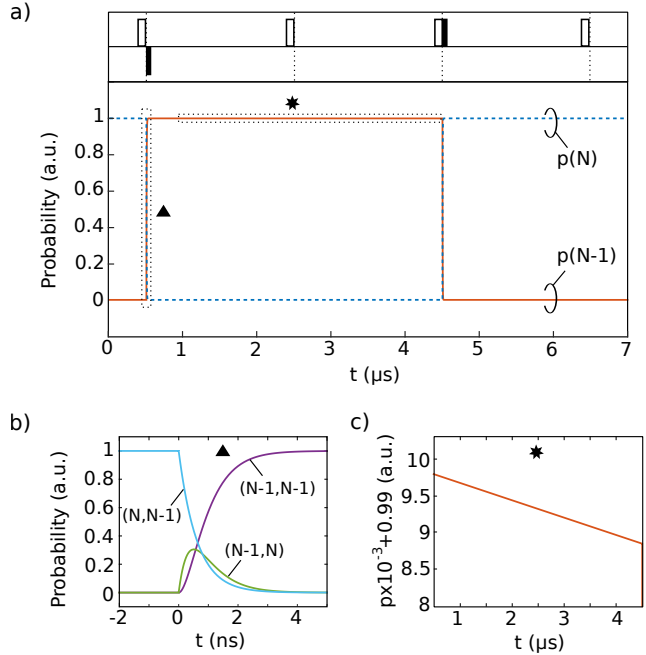

Fig. 7 The response of the double-dot QD-CS device to the sequence of the SET and RESET pulses applied at 0.5 µs and 4.5 µs respectively. Panel b) is a zoom on the joint probability occupation of QD1, $p(N_1) = \sum_{N_2} p(N_1, N_2)$, in the time range of the switching of the voltage pulse, indicated by a triangle in panel a). A zoom on the stability of the metastable state (N-1,N-1) over a µs time scale is shown in panel c), marked by a star in panel a). In the timing diagram in the upper panel, the non filled pulses represent the reading-out pulses, the pulses filled in black represent the SET/RESET pulses

## IV. MODELING QD-CS FINITE-STATE MACHINE OPERATIONS

### A. Definition of SET and RESET operations

Analyzing the transient processes in the QDs, we have shown that one can increase or decrease the switching rates between charge states and the lifetime of metastable states by controlling the time profile of the gate voltage. In order to implement a device operating as a memory, it is necessary to minimize the switching time during the SET and RESET operations and maximize the relaxation time of the metastable charge state during the DoNo operation. We showed that this optimization can be achieved with up to a 6 orders of magnitude ratio in the case of the double dot system. The DoNo operation is nothing else but storage of information during which a QD should keep its charge configuration. During the DoNo operation, the system may be either in the stable state with number of electrons *N* or in the metastable state with number of electrons *N-1*. Therefore, one should ensure that the relaxation time for the metastable state $N-1$ is as long as possible for the value of gate voltage corresponding to DoNo. To fulfill this requirement, we assign the DoNo working voltage of the FSM to point A shown in Figs. 5 and 6. At this voltage the stable charge configuration on QD1 is *N*. The SET voltage pulse drives the system from point A to point B, which lies in the stability region of the charge *N-1* on QD1. In this region the relaxation rate from *N* to *N-1* is large. Staying a finite time at point B, we ensure that QD1 relaxes until it contains *N*-1 electrons. The next jump of the SET voltage pulse is finalized by going back to point A. If the switching from B to A is fast enough with respect to the relaxation time of *N*-1 to *N*, i.e., is diabatic, QD1 remains in the charge state *N*-1, which is metastable at the voltage of point A. The succession of paths A-B-A has thus changed the state of FSM. The relaxation time from the unstable charge state *N-1* to state *N* at the voltage of point A defines the storage time of the single electron dynamic memory. The RESET operation brings QD1 from the *N*-1 charge state to the charge state *N* (see the state diagram in Fig. 1a). Since the charge relaxation time for the transition $(N-1) \rightarrow N$ at point A was designed to be very long, in order to speed up this process and achieve a fast RESET operation, we shift the system into the point C, where the transition rate $\Gamma_{N-1 \rightarrow N}$ is large, and back into the point A. Thus, the RESET operation is defined by the path A-C-A.

B. **Possible implementation of DRAM architecture and Timing**

The ability of the QD to operate as a dynamical FSM with the state diagram shown in Fig. 1 allows us designing a QD single electron dynamic memory that could operate in random access memory architecture as a DRAM cell. In Fig. 8, we show a possible way to interface the DQ-CS device with macroscopic electronics. The device can be considered as a nanoscale analog of a conventional 1T1C DRAM (one Transistor one Capacitor Dynamic Random Access Memory cell, see Fig. 5 b).[23] A small electrical current passing through the single electron transistor which operates as an electrometer, is integrated in time by the capacitor Cstor. When the transistor T1 is open, the capacitor Cstor is discharging through the bit line BL2 providing the output. The output is high when the current through the charge sensor is high enough to charge the capacitor during the time of charge sensing. The gate of the transistor T1 functions as a word line providing access to the memory cell. The SET/RESET voltage pulses are applied to the gate of the double QD through another bit line BL1. The timing diagram shown in the upper panel of Fig. 8 illustrates that the writing and reading memory should be synchronized with proper timing to make the device operating as a memory. A cycle of the device operation consists of the writing phase, the charge sensing phase and the reading out. The time required to switch the SET and RESET pulses shown in the upper panel in Fig. 7a, is 5 ns that is compatible with modern electronics. However to produce the informational output the whole cycle of the device operation needs to include the time required for charge state measurements. Nowadays the fastest electrometers are based on single-electron transistors that provide reliable measurements within several microseconds. We show in Figure 7a that the duration of DoNo operation can reach several µs. Therefore, the dynamical memory FSM could reliably operate at a clock frequency of 0.5 MHz, which provides enough time to accurately read the state of the memory during the DoNo operation. In Fig. 7, we show a FSM operation for three cycles with following sequence of operations: SET-DoNo-RESET. Each cycle is finalized by the pulse applied to the transistor T1 providing reading out. The width of the pulse is related to the time needed to discharge the capacitor Cstor. In modern DRAM systems this time is of several tens of nanoseconds.

The QD-CS device stores its memory during DoNo operations. Each SET or RESET operation leads to memory rewriting. The memory retention time determines the maximum number of DoNo operations can be carried in succession. This number is limited by the relaxation time from the metastable charge state $N-1$ to the charge state $N$. For the parameters used in Fig. 7, the characteristic relaxation time is 2.3 ms. It is the time over which a fit of the probability of the state to an exponential decay drops to 0.37. The state $N-1$ is kept with the probability 0.9 during 0.24 ms, that is 120 clock cycles that includes the reading of the memory state as discussed above.

Although, the state diagram in Fig. 1 a is similar to the state diagram of RS latch,[6] the device is not a conventional synchronous RS latch because the clock input is not explicitly included, however the timing of the pulse is crucial for the device operation.

C. **Bit error rates and possible applications**

The sources of bit errors are: 1) the background charge fluctuation that is always present in the QD devices,[24] 2) the charge sensing process whose output is characterized by a given signal to noise ratio[15] and 3) the relaxation the charge state $N-1$ into charge state $N$ during DoNo operation. According to the simulation results presented above, one cycle of DoNo operation (including the charge state measurement) after the SET operation reduces the occupation probability down to 0.9988 (see Fig. 7 c), which results in a bit error probability of 0.0012. The bit error rate depends on the details of the applied sequence of logic operations: applying DoNo after the RESET operation does not lead to a bit error since the charge state $N$ is stable at the voltage of point A. However applying DoNo after SET increases the bit error. Several hundreds of DoNo operations in sequence almost certainly produce an error in the output.

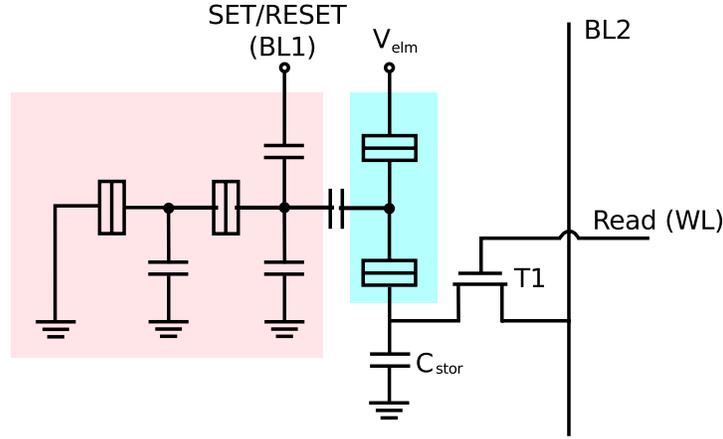

Fig. 8 QD-CS DRAM memory cell

The dependence of the bit error rate on specific logic operations makes the QD-CS device a good candidate for operating in the approximate computing systems[10] as well as for data storage over a single operational cycle. Depending on the sequence of operations the device is characterized by a trade-off between the accuracy of computations and energy consumption: when many combinations of SET-DoNo's are carried out successively the bit error rate can be decreased by reducing the duration of DoNo operation and, consequently, the period of CLK pulses which leads to larger energy consumption. That is consistent with the approximate computing paradigm.

### V. Conclusions

We demonstrated that a system consisting of a chain of QDs and a charge sensor exhibits a dynamic single electron memory based on the control of charge states through pulses of gate voltage. The QD-CS device can therefore operate as a finite state machine. The single electron dynamic memory is based on the asymmetry in the relaxation times of a pair of charge states when the system is driven from one stable state to another. The principle of operation is that the system, stable in a given charge state, is driven out of equilibrium, in a diabatic manner, to a value of voltage where it is not stable. The diabatic switching needs to be faster than the rate of relaxation to the stable state at this voltage. The lifetime of this metastable state is governed by the relaxation rate to the stable state. To implement a memory, one needs that in the direction of inputing, the relaxation rate is fast so that the writing of the memory is fast, while the relaxation time must be very long for memory storing. Although the asymmetry in relaxation rates is present even in the single-QD device, it is significantly enhanced in the chain of quantum dots coupled by tunnel junctions. Our simulations on a chain of two QD's further show that a storage time of several hundreds of microseconds can be achieved for experimentally feasible parameters of the QD. The retention time is then limited by the time required for charge state measurements. The proposed device can be used as a QD DRAM cell for a short-term data storage for performing sequential logic operations, which is particularly relevant for the approximate computing.

**ACKNOWLEDGMENTS**

This work is funded by the FET project TOLOP (318397) of the European Community. FR acknowledges support from the Fonds National de la Recherche Scientifique, Belgium.